\documentclass{sig-alternate}
\pdfoutput=1
\usepackage{common}
\usepackage{url}
\usepackage{subfigure}
\usepackage{verbatim}
\usepackage[pdftex,
            pdfauthor={Derek O'Callaghan, Martin Harrigan, Joe Carthy, Padraig
            Cunningham}, 
            pdftitle={Identifying Discriminating Network Motifs in YouTube
            Spam}, pdfsubject={Identifying Discriminating Network Motifs in
            YouTube Spam}, pdfkeywords={Spam, YouTube, network motif analysis,
            social network analysis}]{hyperref}

\begin{document}

\title{Identifying Discriminating Network Motifs \\in YouTube Spam}

\numberofauthors{1} \author{
\alignauthor Derek O'Callaghan, Martin Harrigan, Joe Carthy, P\'{a}draig Cunningham\\
       \affaddr{School of Computer Science and Informatics}\\
       \affaddr{University College Dublin, Ireland}\\
       \email{\{derek.ocallaghan,martin.harrigan,joe.carthy,padraig.cunningham\}@ucd.ie}}
\date{}

\toappearbox{}

\maketitle

\begin{abstract}
Like other social media websites, YouTube is not immune from the attention of
spammers. In particular, evidence can be found of attempts to attract users to
malicious third-party websites. As this type of spam is often associated with
orchestrated campaigns, it has a discernible network signature, based on
networks derived from comments posted by users to videos. In this paper, we
examine examples of different YouTube spam campaigns of this nature, and use a
feature selection process to identify network motifs that are characteristic of the
corresponding campaign strategies. We demonstrate how these discriminating
motifs can be used as part of a network motif profiling process that tracks the
activity of spam user accounts over time, enabling the process to scale to
larger networks.
\end{abstract}



\section{Introduction}

The popularity and success of YouTube, a video-sharing website consisting for
the most part of user-generated content, continues to grow. Recent
statistics report that it now receives more than four billion video views per
day, with sixty hours of video being uploaded every minute; increases of 30\%
and 25\% respectively over the prior eight
months\footnote{\url{http://youtube-global.blogspot.com/2012/01/holy-nyans-60-hours-per-minute-and-4.html}}.
However, as is the case with many other high-profile social media websites, YouTube has attracted unwelcome attention from
spammers. In this paper, we are concerned with the facility that permits users
to post comments to videos, thus providing a mechanism that can be used to propagate 
spam messages. This can be achieved with
relative ease given the availability of
bots\footnote{\url{http://youtubebot.com/}} that are used to post comments
in large volumes. It is important at this point to distinguish between
\textit{promoters} and \textit{spammers}
\cite{Benevenuto:2009:DSC:1571941.1572047}, with the former using such tools to
encourage channel or video views, whereas the general objective of the latter is
to entice users to visit malicious third-party websites.

\begin{figure}
	\centering
	\includegraphics{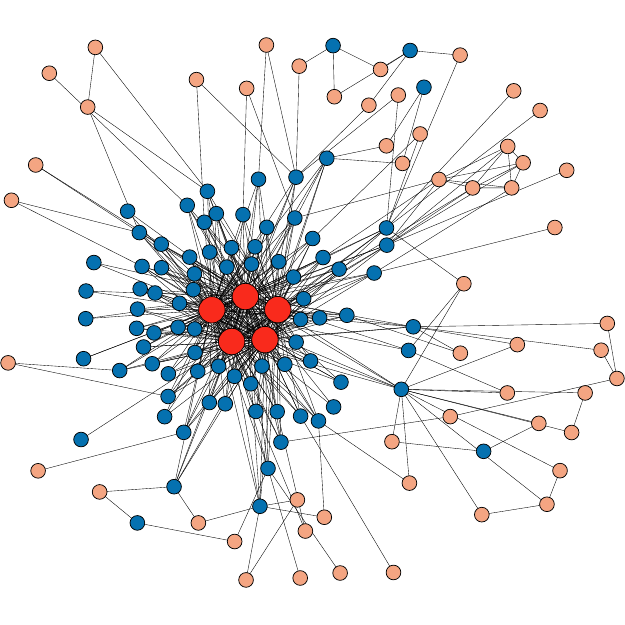}
	\vskip -1.4em
	\caption{Spam campaign targeting YouTube, with a small
	number of accounts each commenting on many videos. Blue nodes are videos,
	orange nodes are user accounts, the five spam accounts are highlighted in red.}
	\label{fig:campaigns}
\end{figure}

One of the findings of our previous work \cite{DBLP:journals/corr/abs-1201-3783}
was the discovery of orchestrated campaigns targeting popular YouTube videos
with bot-posted spam comments. These campaigns are often of a recurring nature,
operating with periodic bursts
\cite{Xie:2008:SBS:1402958.1402979,Gao:2010:DCS:1879141.1879147} of activity on
a continual basis. It appears that different campaign strategies are in use, for
example, a large number of user accounts each commenting on a small number of
videos, or a small number of accounts each commenting on many videos. An example
of a campaign using the latter strategy can be seen in
Figure~\ref{fig:campaigns}, taken from a network derived from comments posted by
users to videos within a particular time period.

Our approach uses the concept of
\textit{network motif
profiling}~\cite{Milo25102002,Milo05032004,Wu:2011:CWP:2065023.2065036} to
detect the recurring activity of these spam campaigns, where motif counts from
the derived networks are tracked over time. Motifs are enumerated on an
\textit{egocentric} basis, where an ego is a single user node within the
networks, and the resulting profiles of motif counts can be used
for user characterization in order to detect spam activity. Our strategy has been to 
consider \emph{all} motifs because we don't assume any prior knowledge of the 
nature of spam campaigns and motifs that might be characteristic of spammers.

However, the
enumeration of all motif instances present in the user networks can
be a lengthy process. At the same time, we have found that certain
discriminating motifs may be used to identify particular strategies and the
associated users as they periodically recur. Given this, the objective of this
paper is to determine from a preliminary analysis a discriminating subset of motifs for profiling, so that the resulting improvement in performance would
permit its application to larger networks.

This paper begins with a description of related work in the domain. Next, the
methodology used by the detection approach is described in detail, from
derivation of the comment-based networks to the subsequent network motif profile
generation. We describe the feature selection process used to identify a set
of discriminating motifs, using profiles associated with known spam user
accounts. Results are then presented from an evaluation of the modified process
where only these motifs are considered. This evaluation focuses on performance
improvement and the detection effectiveness. Finally, the overall conclusions
are discussed, and some suggestions for future work are made.

\section{Related Work}

\subsection{Structural and spam analysis}

The structure of YouTube has been analyzed in a number of separate
studies. Cha~\etal~\cite{Cha:2007:ITY:1298306.1298309} performed an extensive
study that focused on popularity distribution and evolution, content
distribution and the prevalence of duplication and illegal uploads.
Paolillo~\etal~\shortcite{Paolillo08} investigated the social structure with the
generation of a user network based on the friendship relationship, focusing on
the degree distribution. They found that YouTube is similar to other online
social networks with respect to degree distribution, and that a social core
exists between authors (uploaders) of videos. An alternative network based on
related videos was analyzed by Cheng~\etal~\shortcite{Cheng08:4539688}. Given
that the resulting networks were not strongly connected, attention was reserved
for the largest strongly connected components. These components were found to
exhibit \textit{small-world} characteristics~\cite{small-world-watts-strogatz},
with large clustering coefficients and short characteristic path lengths,
indicating the presence of dense clusters of related videos.

Benevenuto~\etal~\shortcite{Benevenuto:2008:UVI:1459359.1459480} created a
directed network based on videos and their associated responses. Similarly, they
found that using the largest strongly connected components was more desirable
due to the large clustering coefficients involved. This was a precursor to
subsequent work concerned with the detection of \textit{spammers} and
content \textit{promoters} within YouTube~\cite{Benevenuto:2008:IVS:1451983.1451996,Benevenuto:2009:DSC:1571941.1572047}.
Features from the video responses networks (e.g. clustering coefficient,
reciprocity) were used as part of a larger set to classify users
accordingly. Other YouTube spam investigations include the recent work of Sureka~\shortcite{DBLP:journals/corr/abs-1103-5044}, based on the detection of
spam within comments posted to videos. A number of features were derived to
analyze the overall activity of users, rather than focusing on individual
comment detection.

An extensive body of work has been dedicated to the analysis of spam within
other online social media sites. For example,
Mishne~\etal~\shortcite{Mishne05blockingblog} suggested an approach for the
detection of link spam within blog comments using the comparison of language
models. Han~\etal~\cite{Han_Ahn_Moon_Jeong_2006} investigated the use of a
collaborative spam filtering scheme to block link spams in blogs.
Gao~\etal~\shortcite{Gao:2010:DCS:1879141.1879147} investigated the
proliferation of spam within Facebook ``wall" messages, with the detection of
spam clusters using networks based on message similarity. This particular study
demonstrated the \textit{bursty} (recurring) and \textit{distributed} aspects of
botnet-driven spam campaigns, as discussed by
Xie~\etal~\shortcite{Xie:2008:SBS:1402958.1402979}. The shortcomings of URL
blacklists for the prevention of spam on Twitter were highlighted by
Grier~\etal~\shortcite{Grier:2010:SUC:1866307.1866311}, where it was found that
blacklist update delays of up to twenty days can occur. This is a particular
problem with the use of shortened URLs, the nature of which was also analyzed by
Chhabra~\etal~\shortcite{Chhabra:2011:PPL:2030376.2030387}. The long-term study
by Lee~\etal~\cite{ICWSM112780} involved the deployment of Twitter honeypots
that resulted in the harvesting of 36,000 candidate content polluters.

\subsection{Network motif analysis}

\textit{Network motifs}~\cite{Milo25102002,ShenOrr02} are structural patterns
in the form of interconnected $n$-node subgraphs that are considered to be
inherent in many varieties of network, such as biological, technological and
sociological networks. They are often used for the comparison of said networks,
and can also indicate certain network characteristics. In particular, the work of
Milo~\etal~\shortcite{Milo05032004} proposed the use of \textit{significance}
profiles based on the motif counts found within networks to enable the
comparison of local structure between networks of different sizes. In this case,
the generation of an ensemble of random networks was required for each
significance profile. An alternative to this
approach~\cite{Wu:2011:CWP:2065023.2065036} involved the
use of motif profiles that did not entail random network generation. Instead,
profiles were created on an \textit{egocentric} basis for the purpose of
characterizing individual \textit{egos}, encompassing the motif counts from the
entirety of egocentric networks within a particular network.

The domain of spam detection has also profited from the use of network motifs or
subgraphs. Within a network built from email addresses~\cite{Boykin:2005:1432647}, a low clustering coefficient (based on the number of
triangle structures within a network) may indicate the presence of spam
addresses, with regular addresses generally forming close-knit communities, i.e.
a relatively higher number of triangles. Becchetti~\etal~\shortcite{Becchetti:2008:ESA:1401890.1401898} made use of the number of
triangles and clustering coefficient as features in the detection of web spam.
These two features were found to rank highly within an overall feature set.
Motifs of size three (triads) have also been used to detect spam comments in
networks generated from blog interaction~\cite{Kamaliha:2008:CNM:1490299.1490781}. It was found that certain motifs were
likely to indicate the presence of spam, based on comparison with corresponding
random network ensembles.

Separately, network motifs have also been used to characterize network traffic~\cite{Allan:2009:UNM:1811982.1812090}. A 
network was created for each application (e.g. HTTP), and
nodes within the network were classified using corresponding motif profiles.


\section{YouTube Comment Networks}

\subsection{Comment processing and network generation}

In our previous work \cite{DBLP:journals/corr/abs-1201-3783}, we described the
collection of a data set\footnote{The data set is
available at \url{http://mlg.ucd.ie/yt}} that was undertaken in order to investigate
contemporary spam comment activity within YouTube. We opted
for a specific selection of the available data given that spam comments in
YouTube tend to be directed towards a subset of the entire video set, i.e. more popular videos generally have a higher probability of attracting attention from spammers, thus
ensuring a larger audience. A decision was made to use only data to which access
was not restricted, namely the comments posted to videos along with the
associated user accounts. To facilitate our objective of analyzing recurring
spam campaigns, we periodically retrieved the details and comments of popular
videos on a continual basis, using the \textit{Most Viewed} standard feed
provided by the YouTube Data
API\footnote{\url{http://code.google.com/apis/youtube/getting\_started.html\#data_api}}.
The API also provides a \textit{spam hint} property within the video comment
meta-data, which is set to \textit{true} if a comment has previously been marked
as spam.
However, this property cannot be considered reliable due to its
occasional inaccuracy, where innocent comments can be marked as spam, while
obvious spam comments are not marked as such. 

Our methodology requires the generation of a network to
represent the comment posting activity of users to a set of videos. Initially, comments made during a
specified time interval are selected from the data set described above. To counteract
obfuscation efforts by spammers in order bypass their detection by any filters, a number of
pre-processing steps must then be executed. During this process, each comment is
converted to a set of tokens, followed by the removal of stopwords along with
any non-Latin-based words, as the focus of this evaluation is English-language
spam comments. Punctuation characters are also removed, and letters are
converted to lowercase. A modified comment text is then generated from the
concatenation of the generated tokens. 

As initial analysis found that spam
comments can often be longer than regular comments, any texts shorter than a
minimum length (currently 25 characters) are removed at this point. Although the
campaign strategies under discussion here are concerned with attracting users to
remote sites through the inclusion of URLs in comment text, comments without
URLs are currently retained. This ensures the option of analyzing other types of
spam campaigns, such as those encouraging channel views, i.e.
\textit{promoters}~\cite{Benevenuto:2009:DSC:1571941.1572047}, along with the
behaviour of regular users.

A network can then be generated from the remaining modified comment texts. This
network consists of two categories of node, \textit{users} and \textit{videos}.
An undirected edge is created between a user and a video if at least one comment
has been posted by the user on the video, where the edge weight represents the
number of comments in question. For the moment, the weight is merely recorded
but is not subsequently used when counting motifs within the network. To capture
the relationship between the users involved in a particular spam campaign,
undirected and unweighted edges are created between user nodes based on the
similarity of their associated comments. Each modified (tokenized) comment
text is converted to a set of hashes using the Rabin-Karp rolling hash
method~\cite{Karp:1987:ERP:1012156.1012171}, with a sliding window length of 3.
A pairwise distance matrix, based on Jaccard distance, can then be generated
from these comment hash sets. For each pairwise comment distance below a
threshold (currently 0.6), an edge is created between the corresponding users if
one does not already exist.

Afterwards, any users whose set of adjacent nodes consists
solely of a single video node are removed. Since these users have commented on
only one video, and are in all likelihood not related to any other users, they are
not considered to be part of any spam campaign. The resulting network tends to
consist of one or more large connected components, with a number of
considerably smaller connected components based on videos with
a relatively minor amount of comment activity. Finally, an approximate
labelling of the user nodes is performed, where users are labelled as spam users if they
posted at least one comment whose \textit{spam hint} property is set to true. All
remaining users are labelled as regular users. Although this can lead to label
inaccuracies, the results shown later in this paper demonstrate that such
inaccuracies will be perceivable.

\subsection{Network motif profiles}

Once the network has been generated, a set of \textit{egocentric} networks can
be extracted. In this context, given that the focus is on user activity, an
\textit{ego} is a user node, where its egocentric network is the induced 
\textit{k-neighbourhood} network consisting of those user and video nodes whose
distance from the ego is at most \textit{k} (currently 2). Motifs from size three to five
within the egocentric networks are then enumerated using FANMOD~\cite{Wernicke01052006}. A set of motif counts is maintained for each ego, where
a count is incremented for each motif instance found by FANMOD that contains the
ego.

A network motif count profile is then created for each ego. As the number of
possible motifs can be relatively large (particularly if directed and/or
weighted edges are considered), the length of this profile will vary for each
network generated from a selection of comment data, rather than relying upon a
profile with a (large) fixed length. For a particular generated network, the
profiles will contain an entry for each of the unique motifs found in the
entirety of its constituent egocentric networks. Any motifs not found for a
particular ego will have a corresponding value of zero in the associated motif
profile.

As mentioned previously, the work of Milo~\etal~\shortcite{Milo05032004}
proposed the generation of a \textit{significance} profile, where the
significance of a particular motif was calculated based on its count in a
network along with that generated by an ensemble of corresponding random
networks. These profiles then permitted the subsequent comparison of different
networks. In this work, the egocentric networks are compared with each other,
and the generation of random ensembles is not performed. An alternative
\textit{ratio} profile $rp$~\cite{Wu:2011:CWP:2065023.2065036} is created for
each ego, where the ratio value for a particular motif is based on the counts
from all of the egocentric networks, i.e.:

\begin{equation}
	rp_i = \frac{nmp_i - \overline{nmp_i}}{nmp_i + \overline{nmp_i} + \epsilon}
\end{equation}

\medskip
Here, $nmp_i$ is the count of the $i^{th}$ motif in the ego's motif profile,
$\overline{nmp_i}$ is the average count of this motif for all motif profiles,
and $\epsilon$ is a small integer that ensures that the ratio is not
misleadingly large when the motif occurs in only a few egocentric networks. To
adjust for scaling, a normalized ratio profile $nrp$ is then created for each
ratio profile $rp$ with:

\begin{equation}
	nrp_i = \frac{rp_i}{\sqrt{\sum{rp_j^2}}}
\end{equation}

\medskip
The generated set of normalized ratio profiles usually contain correlations
between the motifs. Principal components analysis (PCA) is used to adjust for
these, acting as a dimensionality reduction technique in the process.
We can visualize the first two principal components as a starting
point for our analysis. This is demonstrated in Section~\ref{lab:evaluation}.

\section{Discriminating Network Motifs}
\subsection{Identifying Typical Spam Accounts}

In our previous work \cite{DBLP:journals/corr/abs-1201-3783}, we performed an
experiment that tracked two spam campaigns which we had discovered following
manual analysis of the data set. Two distinct campaign strategies were in use,
i.e. a small number of accounts each commenting on many videos (Campaign 1), and
a larger number of accounts each commenting on few videos (Campaign 2). The
experiment was run over a period of seventy-two hours, starting on November
14th, 2011 and ending on November 17th, 2011. In order to track the campaign
activity over time, this period was split into twelve windows of six hours each.
For each of these windows, a network of user and video nodes was derived using
the process described in the previous section. A normalized ratio profile was
generated for each ego (user), based on the motif counts of the corresponding
egocentric network. Principal components analysis (PCA) was then performed on
these profiles to produce 2-dimensional spatializations of the user nodes, using the
first two components. These spatializations acted as the starting point for the
analysis of activity within the twelve time windows.

Although the comment \textit{spam hint} property was used to assist manual
analysis, the previous work was an \textit{unsupervised} exercise in that no
formal labelled training set was generated. In order to determine a set of
discriminating motifs for this paper, a training set with labelled data was
required. For this purpose, we identified time windows containing the highest
number of users belonging to the two campaigns. This was achieved by initially
locating suspicious (outlier) users in the PCA spatializations (e.g.
see Figure \ref{fig:pcacomparison}). In particular, we paid closest attention to
those spatializations where the visible separation was greatest between the
outliers and the majority of regular users.

Having confirmed these users as belonging to the spam campaigns by analyzing
their posted comments, two training sets were created. These sets contained the
normalized ratio profiles for both the respective spam campaign users and the
regular users in the selected time windows. To avoid any interference from
other potential spam campaigns or individual spam users, any other spam users
active in the window (i.e. those who had at least one comment marked as spam)
were removed in both cases. Although the possibility exists for spam users not
to be marked accordingly due to the inconsistency of the \textit{spam hint}
property, it was felt that the percentage of regular users that may in fact have
been spam users would be relatively minor and so no further filtering was
performed.

A third spam campaign (Campaign 3) was also active in this seventy-two hour
period. Its strategy appeared to use a large number of users to each post almost
identical messages to a small ($\sim$1) number of videos. It did not have the
same frequency as the other campaigns (it was not as prevalent in the data set
as the other two), but given the vast number of spam accounts involved it was felt to
be worthy of consideration due to the potential addition of further
discriminating motifs. As before, a third training set containing a set of spam
and regular users for a particular window was created.

\subsection{Feature Selection}

Having generated the labelled training sets, we ranked the features (motifs)
using information gain to determine those that would be characteristic of the classes of interest. 
Due to the nature of the campaigns, the training sets
contained a relatively small number of campaign spam users compared with
that of regular users. To cater for this, we generated random samples from the
training sets by specifying the maximum distribution spread of both classes.
Different ratios of regular users to spam users were available for the three
campaigns, given that varying numbers of users are involved. Those ratios that
generated the highest information gain were chosen. Details of the instance
samples can be found in Table~\ref{tab:infogain}.

\begin{table}[h]
\begin{center}
\begin{tabular}{| l | r | r | r |}
\hline
Campaign & 1 & 2 & 3 \\ \hline\hline
Motifs & 158 & 159 & 158 \\ \hline
Regular:Spam ratio & 5:1 & 3:1 & 2:1 \\ \hline
Regular instances & 30 & 66 & 136 \\ \hline
Spam instances & 6 & 22 & 68 \\ \hline
\end{tabular}
\end{center}
\vskip -1.0em
\caption{Campaign training set samples}
\label{tab:infogain}
\end{table}

\begin{figure}[h!]
	\begin{center}
		\includegraphics{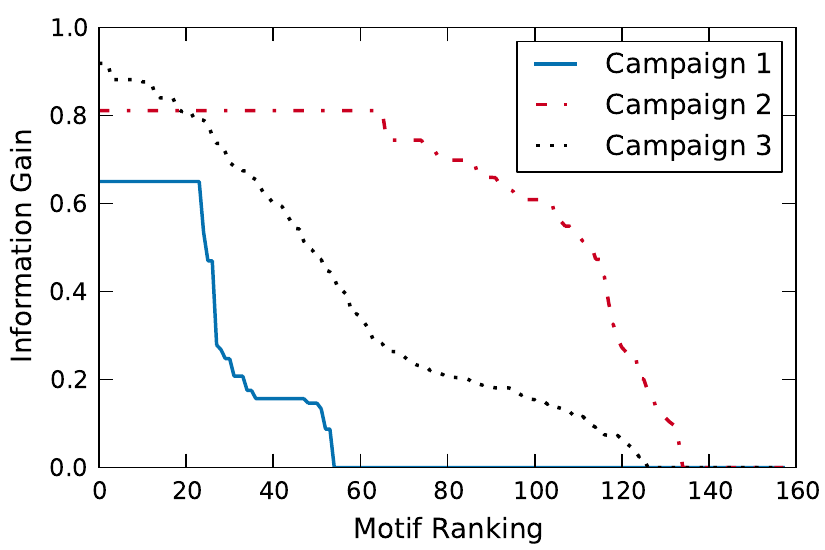}
	\end{center}
	\vskip -1.4em
	\caption{Information gain rankings}
	\label{fig:ig}
\end{figure}

A plot of the motif rankings and corresponding information gain using the
training set samples can be seen in Figure~\ref{fig:ig}.

A selection of 21 motifs was made from those that generated the highest
information gain values, 7 for each of the three campaigns. These motifs can be
found in Table \ref{tab:motifs}. The selection was straightforward for campaign
3, as it was merely a case of selecting the 7 most highly-ranked motifs.
However, as can be seen in Figure~\ref{fig:ig}, a relatively large number of
motifs were found to generate the highest information gain value for both
campaigns 1 and 2. In both cases, we analyzed those motifs ranked highest, and
made a selection based on our knowledge of the campaign strategies following
manual analysis of the data set. For example, campaign 1 uses a small number of
users commenting on a large number of videos, and so those motifs containing
more user-video edges have been selected. For campaign 2, motifs have been
selected that highlight the fact that users appear to be more likely to be
connected to other users rather than videos. The strategy of this campaign is to
employ a larger number of users, each commenting on a small number of videos,
and the potential for connectivity between users is higher given the similarity
of their comments. These motifs also reflect the observed behaviour of users in
the campaign often not commenting on the same videos, as no two users share a
video node neighbour.

\begin{table}[h]
\begin{center}
\begin{tabular}{| c@{}c@{}c@{}c@{}c@{}c@{}c |}
\hline
\multicolumn{7}{|c|}{Campaign~1} \\ \hline
\subfigure{\includegraphics{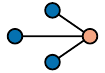}}
&
\subfigure{\includegraphics{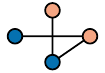}}
&
\subfigure{\includegraphics{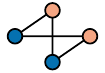}}
&
\subfigure{\includegraphics{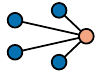}}
&
\subfigure{\includegraphics{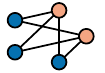}}
&
\subfigure{\includegraphics{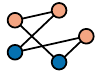}}
&
\subfigure{\includegraphics{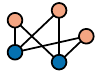}}
\\
1 & 2 & 3 & 4 & 5 & 6 & 7
\\ \hline \hline
\multicolumn{7}{|c|}{Campaign~2} \\ \hline
\subfigure{\includegraphics{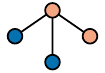}}
&
\subfigure{\includegraphics{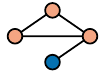}}
&
\subfigure{\includegraphics{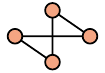}}
&
\subfigure{\includegraphics{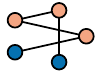}}
&
\subfigure{\includegraphics{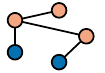}}
&
\subfigure{\includegraphics{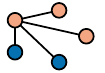}}
&
\subfigure{\includegraphics{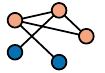}}
\\
8 & 9 & 10 & 11 & 12 & 13 & 14
\\ \hline \hline
\multicolumn{7}{|c|}{Campaign~3} \\ \hline
\subfigure{\includegraphics{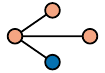}}
&
\subfigure{\includegraphics{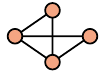}}
&
\subfigure{\includegraphics{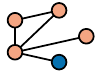}}
&
\subfigure{\includegraphics{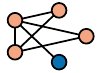}}
&
\subfigure{\includegraphics{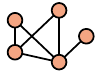}}
&
\subfigure{\includegraphics{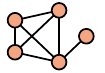}}
&
\subfigure{\includegraphics{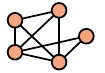}}
\\
15 & 16 & 17 & 18 & 19 & 20 & 21
\\ \hline
\end{tabular}
\end{center}
\vskip -0.8em
\caption{Selected discriminating motifs - blue nodes are videos, orange nodes
are users.}
\label{tab:motifs}
\end{table}
 
Campaign 3 has some similarities with campaign 2, with a large number of
users each commenting on a small number of videos. However, in this case, the
number of users involved is considerably greater, and each user tends to comment
on only one video. The similarity between the comments is also higher, and so
the motifs that were ranked highest are those containing mostly edges between
users, given the number of cliques and near-cliques within the networks
consisting exclusively of user nodes. It should be mentioned here that the use
of such motifs can introduce a limitation to the detection of spam users
associated with such a campaign strategy. In cases where videos have received
highly similar comments from a large number of regular users (e.g. an abundance
of comments containing references to a current holiday or event), the corresponding
egocentric networks may contain many instances of the motifs that are indicative
of this campaign, resulting in their separation in the PCA spatializations.
However, this information may be interesting from a non-spam perspective.

\section{Evaluation}
\label{lab:evaluation}
\subsection{Optimization using discriminating motifs}

Having selected the 21 discriminating motifs, a series of experiments were run
to analyze the spam activity within the collected data set. For the purpose of
this evaluation, the experiments were focused upon tracking any campaigns
having similar strategies to those mentioned in the previous section, i.e.
a small number of accounts each commenting on many videos (Campaign 1), and
variants on a larger number of accounts each commenting on few videos
(Campaigns 2 and 3). The period from December 1st, 2011 to
January 16th, 2012 was split into  windows of six hours each. For each of these
windows, a network of user and video nodes was derived using the process
described earlier, and a normalized ratio profile was generated for each ego
(user), based on the motif counts of the corresponding egocentric network. In
this case however, only the 21 discriminating motifs were enumerated.
Following principal components analysis of these profiles, the spatializations
of the first two principal components act as the starting point for the analysis
of activity within the set of time windows.

As FANMOD \cite{Wernicke01052006} normally enumerates all motifs of a
particular size found within a specified network, a modification to the source
code was required in order to permit a subset of required motifs to also be
specified. We modified the command-line
version\footnote{\url{http://theinf1.informatik.uni-jena.de/~wernicke/motifs/}}
to accept a set of motif identifiers as an additional input parameter. The
enumeration algorithm remains unchanged, instead, the specified motif
identifiers are used to filter the FANMOD output; any instances found for motifs
that are not members of this set are discarded. As we perform
additional post-processing of the enumerated motif instances, this filtering
results in a noticeable improvement in performance.

\begin{table}[h]
\begin{center}
\begin{tabular}{| r | r | r | r |}
\hline
Window & Video nodes & User nodes (spam) & Edges \\ \hline\hline
1 & 322 & 903 (222) & 2974 \\ \hline
2 & 335 & 559 (163) & 1446 \\ \hline
3 & 282 & 745 (157) & 2057 \\ \hline
4 & 283 & 932 (249) & 2475 \\ \hline
5 & 291 & 794 (214) & 2848 \\ \hline
6 & 269 & 470 (123) & 1222 \\ \hline
7 & 281 & 759 (190) & 1931 \\ \hline
8 & 282 & 976 (231) & 2491 \\ \hline
9 & 300 & 689 (163) & 1795 \\ \hline
10 & 330 & 551 (136) & 1701 \\ \hline
11 & 292 & 826 (168) & 2586 \\ \hline
12 & 302 & 958 (225) & 2995 \\ \hline
\end{tabular}
\end{center}
\caption{Network details for all six-hour windows from December 29th, 2011 to
January 1st, 2012.}
\label{tab:windows}
\end{table}

\begin{figure}[h]
	\begin{center}
		\includegraphics{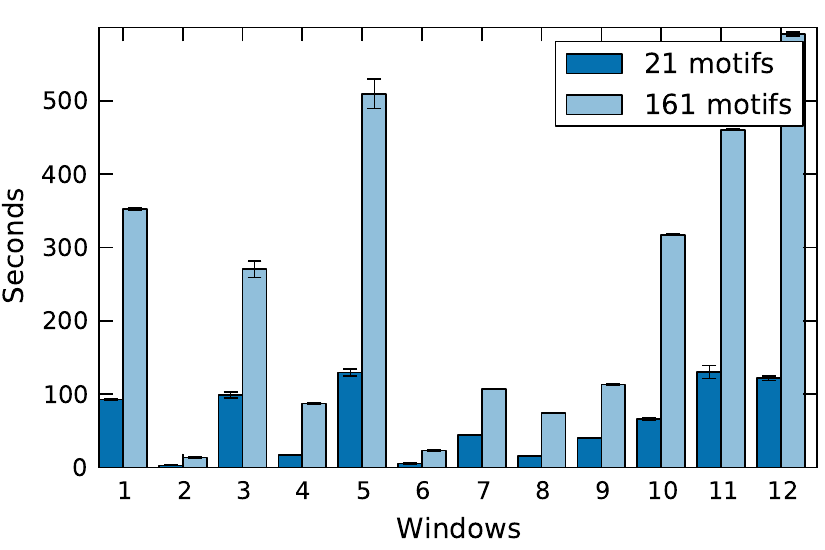}
	\end{center}
	\vskip -.8em
	\caption{FANMOD and associated post-processing execution times for all six-hour
	windows from December 29th, 2011 to January 1st, 2012 (average of 10 runs, error bars indicate the
	standard deviation).}
	\label{fig:timings}
\end{figure}

\begin{figure*}
	\begin{center}
		\mbox{
			\subfigure{\includegraphics{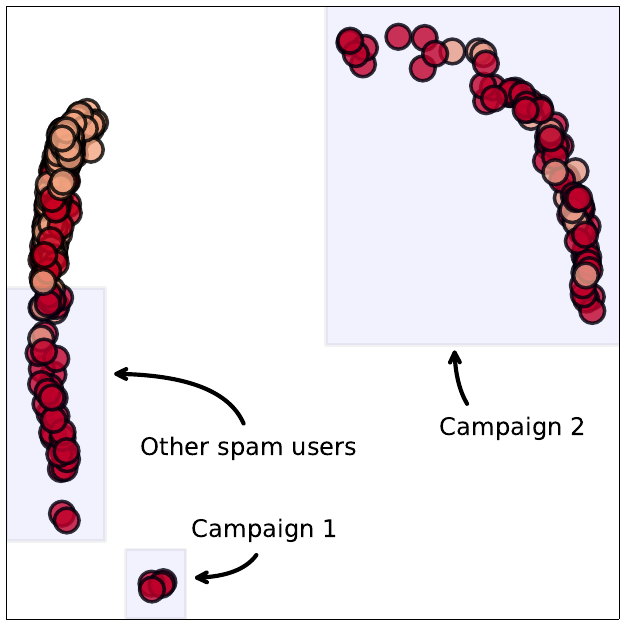}}
			\qquad \qquad
			\subfigure{\includegraphics{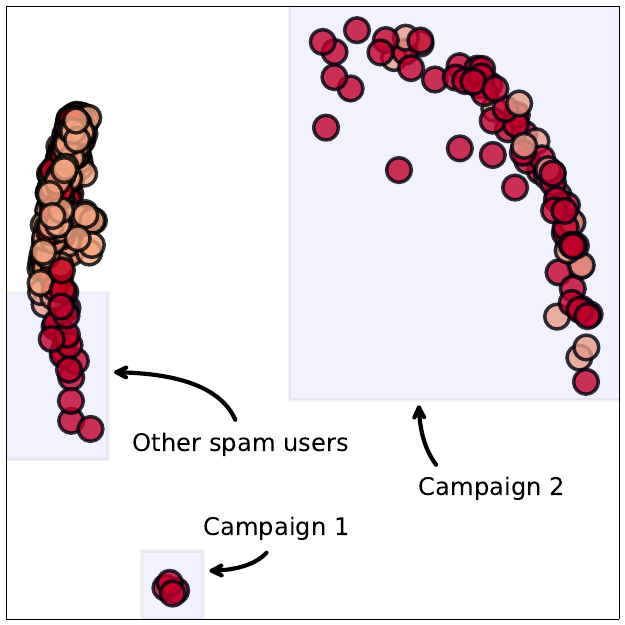}}
			}
	\end{center}
	\vskip -1.4em
	\caption{Spatialization of the first two principal components of the normalized
	ratio profiles for Window 5 (12am to 6am, December 30th, 2011) with 21 motifs
	(left) and all (161) motifs (right). Red nodes are users with comments marked
	as spam, orange nodes are all other users. Most users are regular, and appear
	as overlapping points at the top of the leftmost cluster. Both spam campaigns are
	highlighted.}
	\label{fig:pcacomparison}
\end{figure*}

An analysis of the resulting PCA spatializations found that campaigns 1 and 2
were regularly active during this period. Other spam activity can also be seen,
ranging from an abundance of channel promotion messages from individual users,
to other smaller-scale campaigns. To demonstrate the effect of using a subset of
discriminating motifs, a period of seventy-two hours has been chosen where
campaigns 1 and 2 were active, starting on December 29th, 2011 and ending on
January 1st, 2012. The network details for the twelve six-hour windows can be
found in Table \ref{tab:windows}.

For each of the networks in this period, FANMOD was executed once having
specified the selected 21 motifs, and once without specifying required motifs
(all motifs were enumerated). Figure \ref{fig:timings} contains a plot of the
FANMOD and associated post-processing execution times for each window. As might
be expected, there is a noticeable difference in times between the specification
of the selected motifs, and the case where all motifs are enumerated. On
average, the execution is faster by a factor of 3 when the discriminating motifs
are specified.

\subsection{Effectiveness}

The objective of this section is to demonstrate the effectiveness of the
discriminating motifs in the detection of spam campaigns. As the PCA
spatializations are used as the starting point for any analysis, this will be
addressed through the comparison of two spatializations for Window 5 (12am
to 6am, December 30th, 2011); one created using the normalized ratio profiles
generated from the enumeration of the selected 21 motifs, and the other with
profiles generated from all motifs within the network. These spatializations can
be seen in Figure \ref{fig:pcacomparison}.

At a glance, both spatializations look extremely similar, as expected
given the high correlation between the motifs. Here, users posting at
least one comment marked as spam (using the spam hint property) are in red, all
other users are in orange. The points corresponding to the spam campaign users
have been highlighted accordingly. In both cases, it can be seen that:

\begin{enumerate}
    \item The vast majority of users are regular, and appear as overlapping
    points at the top of the leftmost cluster.
    \item There is a clear distinction between the two different campaign
    strategies, as these points are plotted separately (both from regular users
    and each other). 
    \item The inaccuracy of the \textit{spam hint} comment property is
    demonstrated as the Campaign~2 cluster contains users not coloured in red,
    i.e. none of their comments were marked as spam. Similarly, the reverse is
    true with the large cluster of regular users in that it contains a certain
    number of users coloured in red.
\end{enumerate}

Apart from the highlighted campaign clusters, other spam nodes in the
spatializations have been correctly marked as such. For example, those users
towards the bottom-left appear to be mostly individual spam accounts having
similar behaviour to the strategy of Campaign~1, but on a smaller scale. They
include users encouraging channel views, i.e. \textit{promoters}
\cite{Benevenuto:2009:DSC:1571941.1572047}, and also a number of users belonging
to a separate campaign. Given the similarity of the spatializations, we can
assume that the use of a subset of discriminating motifs in campaign detection
can be just as effective as the results obtained when using all motifs. The
results also confirm the \textit{recurring} nature of these campaigns, given
that the same campaigns were still active over a month after the time period
used in our previous work \cite{DBLP:journals/corr/abs-1201-3783}.

\begin{figure}
	\begin{center}
		\includegraphics{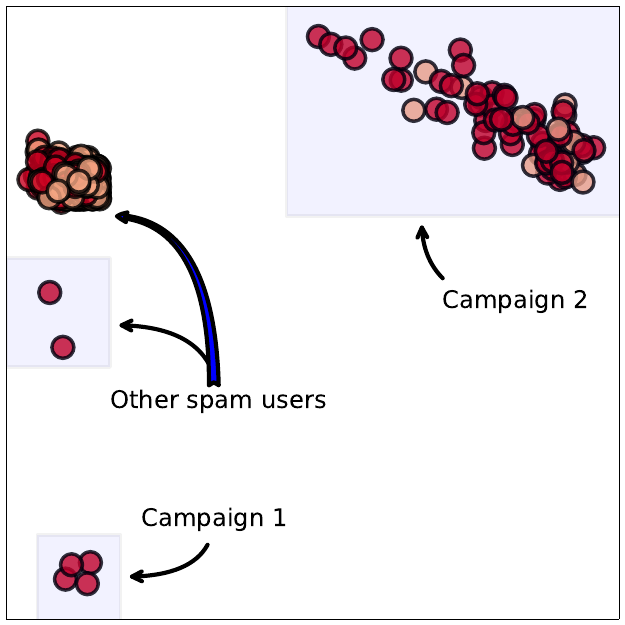}
	\end{center}
	\vskip -1.4em
	\caption{Spatialization of the normalized ratio profiles for Window 5 (12am to
	6am, December 30th, 2011) using motifs 4 and 12 from Table \ref{tab:motifs}
	(same colour coding as Figure~\ref{fig:pcacomparison}). Although both campaigns are separated,
	most of the other spam users have now moved to the cluster of regular users
	in the top-left.}
	\label{fig:pca2motifs}
\end{figure}

With a large number of motifs sharing high information gain values (see
Figure~\ref{fig:ig}), the potential exists for achieving similar
detection using a smaller number of motifs than the set of 21 selected in Table
\ref{tab:motifs}. To illustrate this, we created a final spatialization for
Window 5 that was merely a plot of the normalized ratio values for two
discriminating motifs, one each for campaigns 1 (motif 4) and 2 (motif 12). This
can be seen in Figure \ref{fig:pca2motifs}. As before, both sets of campaign
users are plotted separately from the cluster of regular users and each
other, highlighting the discriminating power of these two motifs. However, some
of the users having behaviour similar to the strategy of Campaign 1 are now
clustered with the regular users. As motif 4 represents a user posting comments
to many videos, it would appear that it alone is not sufficient for the
detection of those users whose behaviour lies somewhere between that of Campaign
1 and regular users.

\section{Conclusions and Future Work}

The presence of orchestrated spam campaigns can be detected in YouTube, where 
sets of spam accounts are used to periodically post comments to popular videos.
These campaigns can employ different strategies, for example, a large number of
user accounts each commenting on a small number of videos, or a small number of
accounts each commenting on many videos. In this paper, having derived networks
from comments posted by users to videos within a set of time periods, we have 
characterized sets of user accounts using a network motif profiling process,
where motifs are enumerated on an egocentric basis. 

Rather than enumerating all motifs within the constituent egocentric networks,
we have found that characterization is also possible with a subset of these
motifs. We have used a feature selection process to select sets of
discriminating motifs associated with campaign strategies within the data set.
Our results from enumerating these particular motifs on networks from later
time periods have demonstrated that their detection of spam user accounts can be
just as effective as that achieved when all motifs are used. At the same time,
the resulting improvement in performance suggests that the use of discriminating
motifs would be appropriate for larger networks, as the enumeration of all
motifs can be a lengthy process. The results also demonstrate the recurring
nature of these campaigns.

In this paper, we have identified three campaign strategies along with their
corresponding discriminating motifs. In the future, it will be useful to
identify other strategies that are present in the data set. For example, we have
found evidence of other campaigns that tend to vary their strategies over time.
This can lead to difficulties in detection, given that their behaviour can
occasionally be similar to that of regular users. Apart from identifying
further discriminating motifs, the current process may need to be modified to
accommodate these strategies. Another possibility for future work is the
derivation of other network representations from the data set, as the comments
network used in this paper is just one abstraction of the underlying network
structure.


\section{Acknowledgments} 
This work is supported by 2CENTRE, the EU funded Cybercrime Centres of Excellence Network and Science Foundation Ireland under grant 08/SRC/I140: Clique: Graph and Network Analysis Cluster.

\bibliography{psosm2012}
\bibliographystyle{abbrv}

\end{document}